\documentclass{aastex631}
\usepackage{amsmath} 

\shorttitle{Last Straw to Break the Equilibrium of a Giant Solar Filament}
\shortauthors{Chen et al.}
\graphicspath{{./}}

\begin{document}

\title{Minifilament Eruptions as the Last Straw to Break the Equilibrium of a Giant Solar Filament}

\correspondingauthor{Hechao Chen, Hui Tian}
\email{hechao.chen@ynu.edu.cn;huitian@pku.edu.cn}

\author[0000-0001-7866-4358]{Hechao Chen}
\affiliation{School of Physics and Astronomy, Yunnan University, Kunming, 650500, China}
\affiliation{School of Earth and Space Sciences, Peking University, Beijing, 100871, China}
\affiliation{Yunnan Key Laboratory of the Solar Physics and Space Science, Kunming, 650216, China}

\author[0000-0002-1369-1758]{Hui Tian}
\affiliation{School of Earth and Space Sciences, Peking University, Beijing, 100871, China}

\author[0000-0003-0565-3206]{Quanhao Zhang}
\affiliation{CAS Key Laboratory of Geospace Environment, Department of Geophysics and Planetary Sciences, University of Science and Technology of China, Hefei, 230026, China}

\author[0000-0001-7693-4908]{Chuan Li}
\affiliation{School of Astronomy and Space Science, Nanjing University, Nanjing 210023, China}
\affiliation{Institute of Science and Technology for Deep Space Exploration, Suzhou Campus, Nanjing University, Suzhou 215163, China}

\author[0000-0002-7153-4304]{Chun Xia}
\affiliation{School of Physics and Astronomy, Yunnan University, Kunming, 650500, China}
\affiliation{Yunnan Key Laboratory of the Solar Physics and Space Science, Kunming, 650216, China}

\author[0000-0003-2686-9153]{Xianyong Bai}
\affiliation{National Astronomical Observatories, Chinese Academy of Sciences, Beijing, 100101, China}

\author[0000-0003-4804-5673]{Zhenyong Hou}
\affiliation{School of Earth and Space Sciences, Peking University, Beijing, 100871, China}

\author[0000-0001-8950-3875]{Kaifan Ji}
\affiliation{Yunnan Observatories, Chinese Academy of Sciences, 396 Yangfangwang, Guandu District, Kunming, 650216, China}

\author{Yuanyong Deng}
\affiliation{National Astronomical Observatories, Chinese Academy of Sciences, Beijing, 100101, China}

\author{Xiao Yang}
\affiliation{National Astronomical Observatories, Chinese Academy of Sciences, Beijing, 100101, China}

\author{Ziyao Hu}
\affiliation{National Astronomical Observatories, Chinese Academy of Sciences, Beijing, 100101, China}


\begin{abstract}

Filament eruptions are magnetically driven violent explosions commonly observed on the Sun and late-type stars, sometimes leading to monster coronal mass ejections that directly affect the nearby planets' environments. More than a century of research on solar filaments suggests that the slow evolution of photospheric magnetic fields plays a decisive role in initiating filament eruptions, but the underlying mechanism remains unclear. Using high-resolution observations from the \textit{Chinese H$\alpha$ Solar Explorer}, the \textit{Solar Upper Transition Region Imager}, and the \textit{Solar Dynamics Observatory}, we present direct evidence that a giant solar filament eruption is triggered by a series of minifilament eruptions occurring beneath it. These minifilaments, which are homologous to the giant filament but on a smaller tempo-spatial scale, sequently form and erupt due to extremely weak mutual flux disappearance of opposite-polarity photospheric magnetic fields. Through multi-fold magnetic interactions, these erupting minifilaments act as the last straw to break the force balance of the overlying giant filament and initiate its ultimate eruption. The results unveil a possible novel pathway for small-scale magnetic activities near the stellar surface to initiate spectacular filament eruptions, and provide new insight into the magnetic coupling of filament eruptions across different tempo-spatial scales.
\end{abstract}



\section{Introduction} \label{sec:intro}

Filaments are one type of magnetized plasma structures commonly observed in the coronae of solar-like and later-type stars  \citep{1996IAUS..176..449C}. These unique phenomena are the key to uncovering the intricate interplay between astrophysical magnetic fields and plasmas in the universe. Filaments, also known as stellar prominences when appearing above the stars' limb, usually come to an end through magnetically driven eruptions. Once erupting, they are most likely accompanied by flares and spectacular coronal mass ejections (CMEs) with a rapid magnetic energy release \citep{2011LRSP....8....1C,2011LRSP....8....6S,2013AdSpR..51.1967S}, which may adversely affect the nearby orbiting planets and even pose a threat to their habitability  \citep{2020IJAsB..19..136A,2023ScSnT..53.2021T}. 

On the Sun, filaments have been studied  for over a century with spatially-resolved observations due to their proximity \citep{2014LRSP...11....1P}. Extensive solar observations have shown that filaments form and erupt from the sunspot belts to the polar crown \citep{2004ApJ...614.1054J,2012ApJ...757..168S}, with timescales ranging from minutes to months and lengths spanning from $\sim10^{3}$ km up to $10^{5}$ km  \citep{2000ApJ...530.1071W,2011ApJ...738L..20H,2017ApJ...844..131P,2020ApJ...902....8C}. Their eruptions exhibit a very broad energy releasing range ($10^{25}$ to $10^{32}$ erg) and a wide mass ejection range ($10^{12}$ to $10^{15}$ g)  \citep{2023ApJ...943..143K}. The scale-free self-similar nature of magnetohydrodynamics  (MHD)  suggests that these multi-scale filaments are cross-scale analogs that form via a common mechanism \citep{2020ApJ...902....8C}. Moreover, a few studies state that solar eruptive activities driven by filaments at different scales, including stellar-scale CMEs and small-scale coronal jets, may also be triggered by a similar mechanism \citep{2010ApJ...718..981R,2010ApJ...710.1480S,2015Natur.523..437S,2017Natur.544..452W},
but the conditions that make filament eruptions inevitable remain a subject of active debate \citep{2014IAUS..300..184A,2021NatAs...5.1096A}. 
Observations and numerical simulations of filament eruptions over the past decades have suggested that the slow evolution of photospheric magnetic fields plays a decisive role in initiating filament eruptions. 
Many possible mechanisms, i.e., breakout \citep{1999ApJ...510..485A,2017Natur.544..452W} and tether-cutting reconnection \citep{2001ApJ...552..833M,2021NatAs...5.1126J}, as well as ideal MHD instabilities \citep{1991ApJ...373..294F,2004A&A...413L..27T,2006PhRvL..96y5002K}, have been proposed to explain the initiation of filament eruptions. However, all of these models concentrate exclusively on the eruption of individual filaments within specific spatial-temporal scales, which prevents us from uncovering the potential physical connections between filament eruptions across different spatial-temporal scales.

Here, we report the discovery of a new type of physical coupling between miniature and giant filament eruptions along the same polarity inversion line (PIL), which has been never been considered in current theories and observations. Minifilament eruptions driven by extremely weak photospheric flux cancellation are found to act as the last straw to break the initial equilibrium of an overlying giant filament through multi-fold magnetic interactions. This finding represents the first solid evidence that large-scale filament eruptions could be triggered by similar eruptions at much smaller scales, opening a new window to understand the physical connection among different filament eruptions across different tempo-spatial scales. Moreover, this unique coupling-eruption pattern is found to be driven by a non-uniform photospheric flux cancellation process. By incorporating the non-uniform characteristics of actual photospheric magnetic flux cancellation, we successfully interpret all crucial features of this unique coupling-eruption pattern of miniature and giant filaments into a unified physical framework, and add new and significant elements into the traditional flux-cancellation model \citep{1989ApJ...343..971V}.

\section{Instruments}

The giant filament eruption of interest occurred near the southwest limb of the solar disk. It caused a C-class flare and notable coronal dimming in its source region, as well as an accompanying CME in the higher corona. Several space-borne observatories, including the \textit{Chinese H$\alpha$ Solar Explorer} \citep[CHASE,][]{2022SCPMA..6589602L}, the \textit{Solar Upper Transition Region Imager} \citep[SUTRI,][]{2023RAA....23f5014B},  and the \textit{Solar Dynamics Observatory} \citep[SDO,][]{2012SoPh..275....3P}, as well as the \textit{Solar TErrestrial RElations Observatory} \citep[STEREO,][]{2008SSRv..136...67H}, have well recorded the activation and eruption of this event. This allows us to investigate the triggering process of the giant filament in details.

The SUTRI is China’s first telescope to probe the solar transition region and is one of the payloads onboard the first spacecraft of the Space Advanced Technology demonstration satellite series, which was launched in July 2022. As a low-cost experiment, SUTRI aims to test the on-orbit performance of EUV CMOS imaging camera and Sc/Si multi-layer reflecting mirror. SUTRI observes the full solar disk and samples plasma at a temperature of 0.5 MK \citep{2017RAA....17..110T} using the Ne VII 465 \AA~line, with a time cadence of 30 s and a spatial resolution of $\sim$8$^{\prime\prime}$. This unique passband has rarely been used by past and existing solar extreme ultraviolet imagers. In recent years, SUTRI has contributed to research on solar eruptive events, achieving a wealth of research results \citep[e.g.,][]{2023A&A...675A.169L,2023ApJ...949...66L,2023ApJ...953..148S,2023ApJ...959...69H,2024A&A...687A.190H,2024MNRAS.533.3255Z,2024ApJ...968...85Z}.
The filament eruption under our study was one of the most interesting events captured by SUTRI after it started routine observations on 2022 September 4. SUTRI Level-0.9 465  \AA~images were used to investigate the dynamics of this filament eruption.
 
The CHASE, dubbed as ``Xihe" in Chinese (the Goddess of the Sun in China), was launched into a Sun-synchronous orbit on October 14, 2021. As the scientific payload of CHASE, the H$\alpha$ Imaging Spectrograph (HIS), for the first time, acquires seeing-free H$\alpha$ solar spectroscopic observation with high temporal and spectral resolutions. The CHAE/HIS performs its routine observations in two modes: raster scanning mode (RSM) and continuum imaging mode. In our study, the RSM Level-1 H$\alpha$  (6559.7$-$6565.9 \AA) spectra under the full-Sun scanning mode were used to investigate the dynamics of the filament eruption and its precursor activity, with a temporal resolution of  about 1 min and a pixel spectral resolution of 0.024 \AA. The Level 1 RSM spectra were produced from the Level 0 data via a detailed calibration procedure \citep{2022SCPMA..6589603Q}, which include dark-field and flat-field correction, slit image curvature correction, wavelength and intensity calibration, as well as coordinate transformation. The CHAE/HIS  H$\alpha$ spectra were utilized to reconstruct two-dimensional Doppler maps using a weight-inverse-intensity method. To eliminate the effects of solar rotation and satellite motion, a reference line profile was obtained by averaging spectra from a nearby quiet region. This reference line profile was then employed for rest wavelength calibration. It is important to note that the CHASE observational data for this event contains some time gaps. Therefore, GONG H-alpha data was used as a supplement to construct the time-distance plot.


The Atmospheric Imaging Assembly \citep[AIA,][]{2012SoPh..275...17L} on board the SDO observes the full solar disk in 10 EUV and UV passbands, with a spatial pixel size of $\sim$0.6$^{\prime\prime}$ and a time cadences of 12 s in EUV passbands (24 s in UV passbands). The Helioseismic and Magnetic Imager \citep[HMI,][]{2012SoPh..275..207S} on board the SDO measures the full-disk solar photospheric magnetic field at six wavelength positions across the Fe I 6173 \AA~spectral line. Here, AIA images in 304 \AA, 171 \AA, and 193 \AA~passbands were used to investigate the plasma dynamics during the filament eruption, and the HMI light-of-sight (LOS) magnetograms were employed to inspect the photospheric magnetic field evolution beneath the eruption.  The SUTRI and SDO data were reduced using the standard analysis tool available in the SolarSoftWare IDL (SSWIDL) package. To compensate for the solar rotation, all the SDO and SUTRI images taken at different times are aligned to an appropriate reference time.  

\section{Observations and Results} \label{sec:results}
\subsection{Overview}
According to the long-term CHASE H$\alpha$ and HMI LOS magnetogram monitor (Figure~\ref{figA1}), this giant quiescent filament appeared in an unnamed active region remnant since 11 September  2022. It exhibited prominent barbs and an apparent length exceeding 300 Mm. From September 8 to 16, long-term photospheric lux cancellation persistently occurred at the core of the active region remnant along the PIL of the giant filament. This cancellation exhibited a gradual reduction in the spatial scale, shrinking from the initial active region size to a few Mm. In this course, the giant filament remained relatively stable at first several days, and its middle segment became highly diffuse in H$\alpha$ images starting from 13 September. The giant filament eventually erupted at approximately 20:00 UT on 15 September. Before its final eruption, a series of preceding small-scale eruptive activities occurring below its middle segment during 15:00 to 19:00 UT (see the associated movie in Figure 1). At first sight, these small-scale eruptions appeared to disturb the magnetic fields of the overlying giant filament, potentially triggering its subsequent eruption.

\begin{figure}[ht!]
\plotone{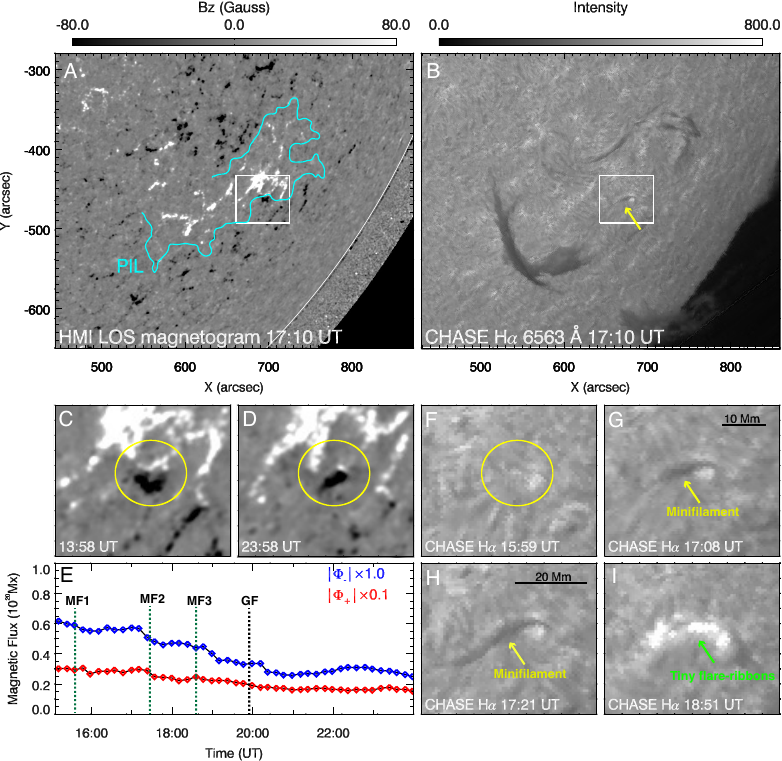}
\caption{The formation and eruption of minifilaments underneath a giant filament due to small-scale flux cancellation on September 15th.  (A) The SDO/HMI LOS magnetogram at 17:10 UT. The cyan solid line denotes the polarity inversion line (PIL) where both the minifilaments and giant filament form. (B) The CHASE H$\alpha$ image at 6562.8 Å. (C-D) The convergence and cancellation of opposite-polarity small-scale magnetic flux concentrations. Another unrelated prominence is observed at the limb (x = 730, y = -600) and remains stable during the giant filament eruption. The field of view (FOV) is indicated by the white rectangle in (A). E) Evolution of magnetic flux during 15:00$-$24:00 UT within the field of view (FOV) of (C). Four vertical dashed lines indicate the eruption times of three minifilaments (abbreviated as MF1, MF2, and MF3) and the overlying giant filament (abbreviated as GF), respectively. (F-I) Snapshots of minifilament activity in the FOV of (C). Features of interest are marked in each panel; see the text for details. 
An animation of panel (B) is available in the online journal. This animation presents a time sequence of 720 s LOS magnetograms and AIA 193 \AA~images spanning from 15:34 UT to 23:58 UT on 15 September 2022, with a total duration of 10 s.
}
\label{fig1}
\end{figure}

\subsection{Minifilaments and extremely weak flux cancellation beneath the giant filament}
Thanks to the unprecedented seeing-free H$\alpha$  observation by CHASE, we clearly see that these small-scale eruptive activities below the pre-eruption giant filament are a sequence of homologous minifilament eruptions. In the CHASE H$\alpha$  image taken at 17:08 UT, it is clear that one minifilament existed right below the middle section of the pre-eruption giant filament (Figure~\ref{fig1}(B)). The simultaneous HMI LOS magnetogram shows that this minifilament and the giant filament resided along the same filament channel (or PIL) (Figure~\ref{fig1}(A)). This unique magnetic configuration of co-existing mini- and giant filaments is reminiscent of the reported double-decker magnetic flux rope system in active region filaments \citep{2012ApJ...756...59L}. 
Based on filament chirality rules, we find that the giant filament should be supported in a magnetic flux rope (MFR) configuration with a sinistral chirality (Methods in Appendix \ref{sec:C}). The nonlinear force free field extrapolations based on the vector photospheric magnetic measurements reveal that the minifilament also resided in a sinistral small-scale MFR (Figure~\ref{figA1}; Methods in Appendix \ref{sec:C}).

The photospheric source region of this minifilament is a small-scale mixed-polarity region, locating at the middle section of the whole curved PIL (see the white box in  Figure~\ref{fig1} (A-B)). In this region, opposite-polarity magnetic flux concentrations underwent a very slow and long-lasting convergence and cancellation at a very short section of the PIL on September 15 (yellow circles in  Figure~\ref{fig1} (C and D) and associated movie). The magnetic flux cancellation refers to mutual approaching and subsequent disappearance of opposite-polarity magnetic fragments in the photosphere \citep[see references in][]{1987ARA&A..25...83Z}.  
From 14:00 to 21:00 UT on September 15, the total reduction of negative flux is $\sim$3.1$\times10^{19}$ Mx and the rate of flux cancellation is $\sim$2.6$\times10^{18}$ Mx per hour on average.  The boundary of canceling opposite magnetic polarities provides a necessary environment for the buildup of filament magnetic fields \citep{1998SoPh..182..107M,2010SSRv..151..333M}.
During this extremely weak flux cancellation, three homologous minifilaments sequently formed within 10$-$30 minutes, and erupted above this canceling site at 15:30 UT, 17:30 UT, and 18:40 UT, respectively (refer to the eruption times of these minifilaments that marked in Figure~\ref{fig1}(E)).
Assuming that equal amount of positive and negative flux cancels and all canceled flux builds the minifilament magnetic field within ten to thirty minutes, we can roughly estimate that$\sim$  {(0.87--2.6)} $\times10^{18}$ Mx of magnetic flux contributes to the buildup of each minifilament on average. This kind of canceled flux is a minuscule fraction of the total magnetic flux of the whole giant filament channel (about $\sim$6.8$\times10^{21}$ Mx). The CHASE observation covered the formation of the second minifilament (see Figure~\ref{fig1} (F to H)) and the eruption of the third one (Figure~\ref{fig1} (I)). As shown in Figure~\ref{fig1} (G to H), the second minifilament had a projected length of about 10 Mm at the very beginning of its birth at 17:08 UT, and soon it grew to 20 Mm near its eruption at 17:21 UT. Though these minifilaments are very short-lived features formed and erupted on a time scale of only tens of minutes, their eruptions continuously disturbed the overlying giant filament (see Figure~\ref{fig2} and Figure~\ref{fig3} (B and C)).

\subsection{Activation of the giant filament due to minifilament eruptions}
\begin{figure}[ht!]
\plotone{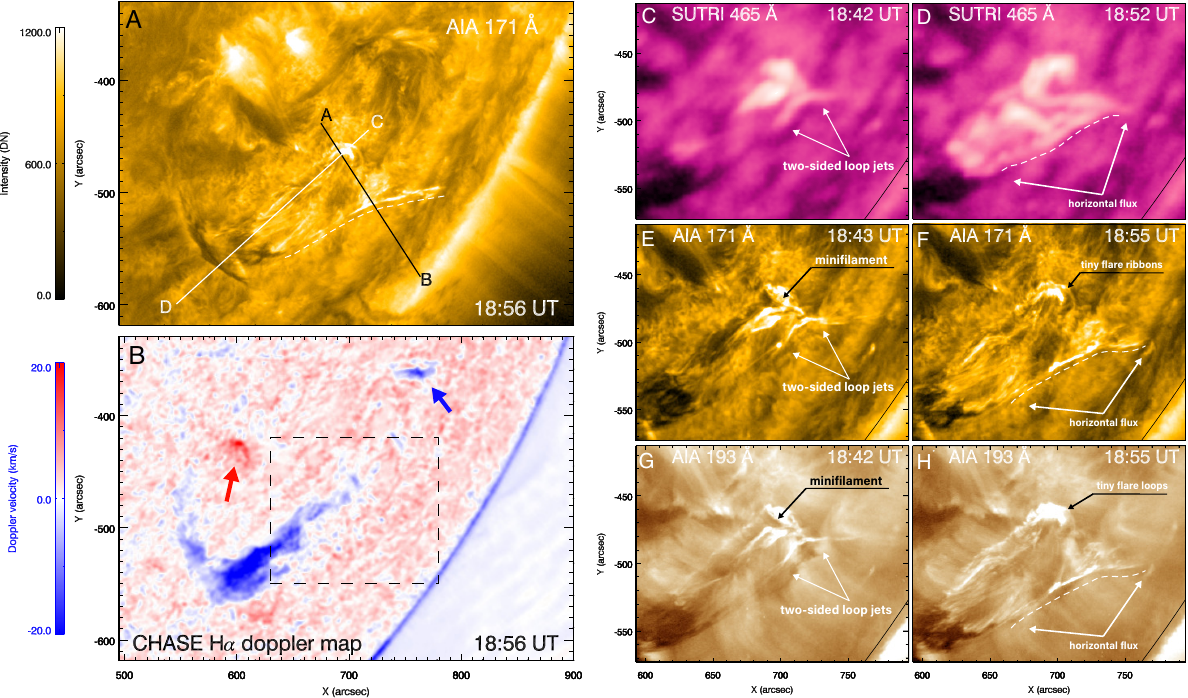}
\caption{ Magnetic interaction between the erupting minifilament and the overlying giant filament during the period from 18:30 UT to 19:30 UT on 2022 Sep 15. (A) The AIA 171 Å image at 18:56 UT. The solid lines AB (in black) and CD (in white) indicate the slice positions for the two space-time diagrams in Fig.4, respectively. The white dashed line outlines a bundle of upward-moving horizontal magnetic field or flux. (B) The Doppler map reconstructed from the CHASE H$\alpha$ spectroscopic observation at 18:56 UT. Red/Blue arrow marks the redshift/blueshift at the left/right feet of the erupting giant filament, respectively. The black dashed rectangle represents the FOV of (C-H). (C-H) Zoom-in snapshots of the magnetic interaction process in SUTRI 465 Å, AIA 171 Å, and AIA 193 Å images. Features of interest are marked in each panel; see the text for details. An animation of this figure is available in the online journal, showing the interaction between the erupting minifilament and the giant filament. The duration of the animation is 3 s.}
\label{fig2}
\end{figure}

Solid evidence of magnetic interaction occurring between the third erupting minifilament and the giant filament were shown in Figure~\ref{fig2} and associated Movie. This minifilament eruption occurred at about 18:40 UT and caused tiny chromospheric flare ribbons (Figure~\ref{fig1}(I) and Figure~\ref{fig2}(F)). At about 18:42 UT, the erupting minifilament structure lifted up and directly reconnected with the overlying giant filament fields. As a result of this reconnection, their interaction site soon radiated enhanced EUV emission, which can be simultaneously detected in almost all the AIA EUV passbands and the SUTRI 465 \AA~passband (Figure~\ref{fig2} (C, E, and G)). At the same time, a pair of roughly antiparallel plasma beams soon developed from the central emission region, forming a series of so-called two-sided loop jets or bidirectional jets \citep[e.g.,][]{2013ApJ...775..132J,2018ApJ...861..108Z,2019ApJ...871..220S,2019ApJ...883..104S,2023MNRAS.520.3080T}. The bidirectional jets occurred at the interface between the overlying giant filament and the erupting minifilament, $\sim$ 24 Mm above the photosphere (see Figure~\ref{fig3}(D); Methods in Appendix \ref{sec:E}).  The projected velocities of the bidirectional jets were measured using a time-distance analysis and found to be $\sim$ 134 km s$^{-1}$ and 54 km s$^{-1}$, respectively. The higher southward velocity is likely due to a less pronounced projection effect.

Several minutes later, a series of newly-formed horizontal bright thread-like plasma structures were developed and then detached from the bidirectional jet spires, which can be clearly observed in both  171 \AA~and 193 \AA, as well as 465 \AA~passbands (see Figure~\ref{fig2} (A, D, F, and G) and associated movie). Later on, they rapidly ascended upward and eventually merged into the overlying giant filament, with a projected velocity of $\sim$ 60 km s$^{-1}$ (Figure~\ref{fig3}(A)). 
Considering the high magnetic Reynolds number of the hot corona, the plasma is frozen to the magnetic field; so these ascendent thread-like plasma structures very likely trace out a series of upward-transporting horizontal magnetic field from the minifilament eruption. Owing to the horizontal magnetic field merging into the overlying giant filament structure and the mass ejection from bidirectional jets, enhanced mass flows immediately appeared as evident blueshifts along the giant filament axis in the simultaneous CHASE H$\alpha$ Doppler map (Figure~\ref{fig2} (B) and associated movie). 
At the two ends of the giant filament, enhanced mass flows caused patchy blueshift and redshift signals along the line of sight, respectively. Due to enhanced mass flows, possible mass drainage may facilitate the lift of the giant filament by reducing its filament weight  \citep[e.g.,][]{2018ApJ...862...54F}. After this sudden activation process, the giant filament soon stepped into its eruption phase at about 20:00 UT.

\subsection{The last straw to break the quasi-static equilibrium of the giant filament}

\begin{figure}[ht!]
\plotone{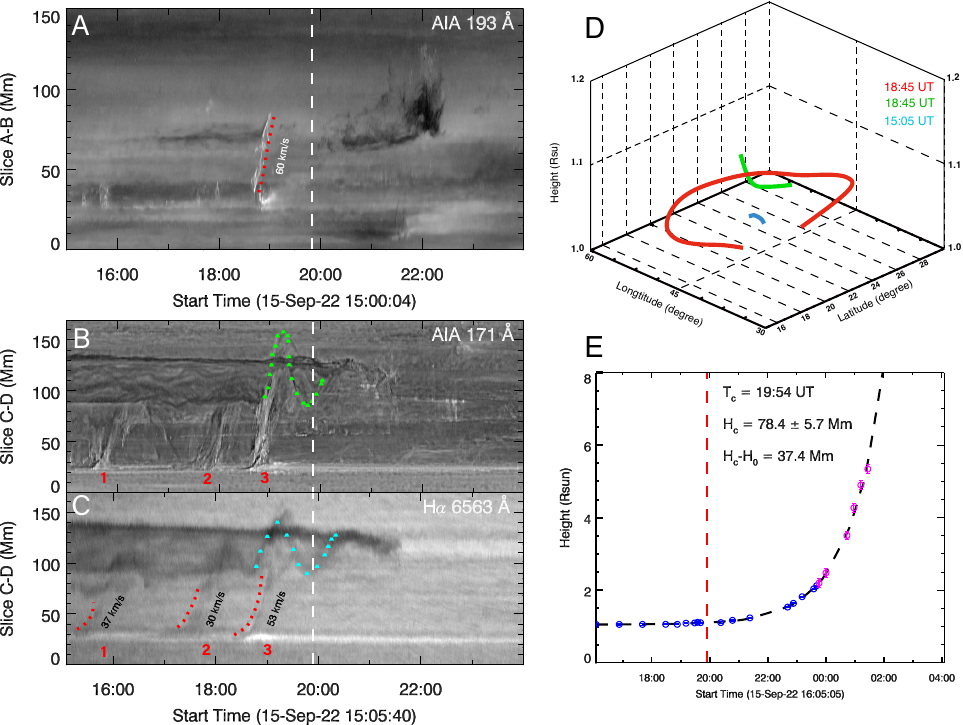}
\caption{Eruptive dynamics and three-dimensional (3D) reconstruction of the giant filament. (A) Space-time diagram of AIA 193 \AA~ images along slice AB in Fig 2A. The red dotted line highlights an upward-moving horizontal flux feature and its moving velocity is around 60 km s$^{-1}$. (B-C) Space-time diagrams of AIA 171  \AA~  and GONG H$\alpha$ images along slice CD in Fig 2A, respectively. Three minifilament eruptions are marked by red numbers (1, 2, 3), respectively. Their velocity are fitted to be 37 km s$^{-1}$, 30 km s$^{-1}$, and 53 km s$^{-1}$, respectively. The green/cyan diamonds highlight the plasma oscillation (amplitude $\sim$ 50-70 Mm) of the giant filament near its south part. (D) The 3D morphology of minifilament (blue), giant filament (red), and two-sided loop jets (green) reconstructed through stereoscopic observations of STEREO-A/EUVI and SDO/AIA in 304  \AA~passbands. (E) Height of the main axis top of the erupting giant filament as a function of time and its fitting result (black dashed line). The blue dots represent stereoscopic measurements from SDO/AIA and STEREO/EUVI, and the purple dots are measured from LASCO/C2. The white dashed lines in (A-C) and the red dashed line in (E) indicate the fitted onset time (19:54 UT) of the giant filament eruption.}
\label{fig3}
\end{figure}

With the aid of three-dimensional (3D) stereoscopic reconstruction (see Figure~\ref{figA6} and methods in Appendix \ref{sec:E}), we find that the height-time profile of the giant filament consists of a slow-rise phase and a rapid-acceleration phase (Figure~\ref{fig3} (E)). This profile can be well characterized by a fitting function consisting of linear and exponential components (Fitting methods in Appendix \ref{sec:E}). The fitting result reveals that (1) the slow-rise phase of the giant filament began at $\sim$16:00 UT (which coincided with the first eruption of the minifilament) and its ascent velocity is quite slow ($\sim$ 4$-$6 km s$^{-1}$), indicating a quasi-static evolution; (2) the giant filament suddenly erupted upward with a velocity of $\sim$140  km s$^{-1}$ and the onset time of the rapid-acceleration phase is about 19:54 UT. This acceleration process was accompanied by a solar flare, showing up as a simultaneous AIA 171 \AA~flux enhancement in its coronal source region (Figure~\ref{figA7} (E) and associated movie). These results suggest that the giant filament eruption was related to a quasi-static evolution and then a flare-related energy releasing process \citep[e.g.,][]{2020ApJ...894...85C}. 

To understand the abrupt kinematic transition from the quasi-static evolution to the rapid acceleration, we present time-distance diagrams along slice AB and CD in Figure~\ref{fig2} (A). These diagrams allow a quantitative investigation for the dynamic response of the giant filament to horizontal magnetic field merging and mass ejections from multiple minifilament eruptions, respectively. The time-distance diagram along slice AB in AIA 193  \AA~images (Figure~\ref{fig3} (A)) clearly shows that during the third minifilament eruption, obvious horizontal flux moving along with the erupting minifilament merged into the overlying giant filament. The time-distance diagrams along slice CD in AIA 171 \AA~and H$\alpha$ images (Figure~\ref{fig3} (B and C)) consistently reveal that due to the mass ejections from three subsequent minifilament eruptions, the giant filament was remarkably disturbed and its cross-section underwent an apparent compression. The third minifilament eruption, as the strongest one with a projected eruption velocity of 53 km s$^{-1}$, caused a very prominent oscillation of the giant filament. This large-amplitude oscillation lasted for one cycle and its amplitude is $\sim$ 50 Mm in AIA 171 \AA~images and 72 Mm in H$\alpha$ images, respectively. Such large-amplitude oscillations are often considered to be possible precursor signals for the imminent filament eruptions \citep[e.g.,][]{2014ApJ...795..130S,2023ApJ...959...71D,2024MNRAS.533.3255Z,2024ApJ...963..140Z}. After the horizontal magnetic field merging into the overlying giant filament structure and the appearance of prominent filament oscillation, this giant filament abruptly entered its accelerating eruption phase from a quasi-static evolution phase. Based on the above observations and kinematic analysis, we conclude that a series of minifilament eruptions act as the last straw to break the quasi-static equilibrium of the giant filament, facilitating its final eruption.

\subsection{Non-uniform decay of background magnetic fields}

\begin{figure}[ht!]
\plotone{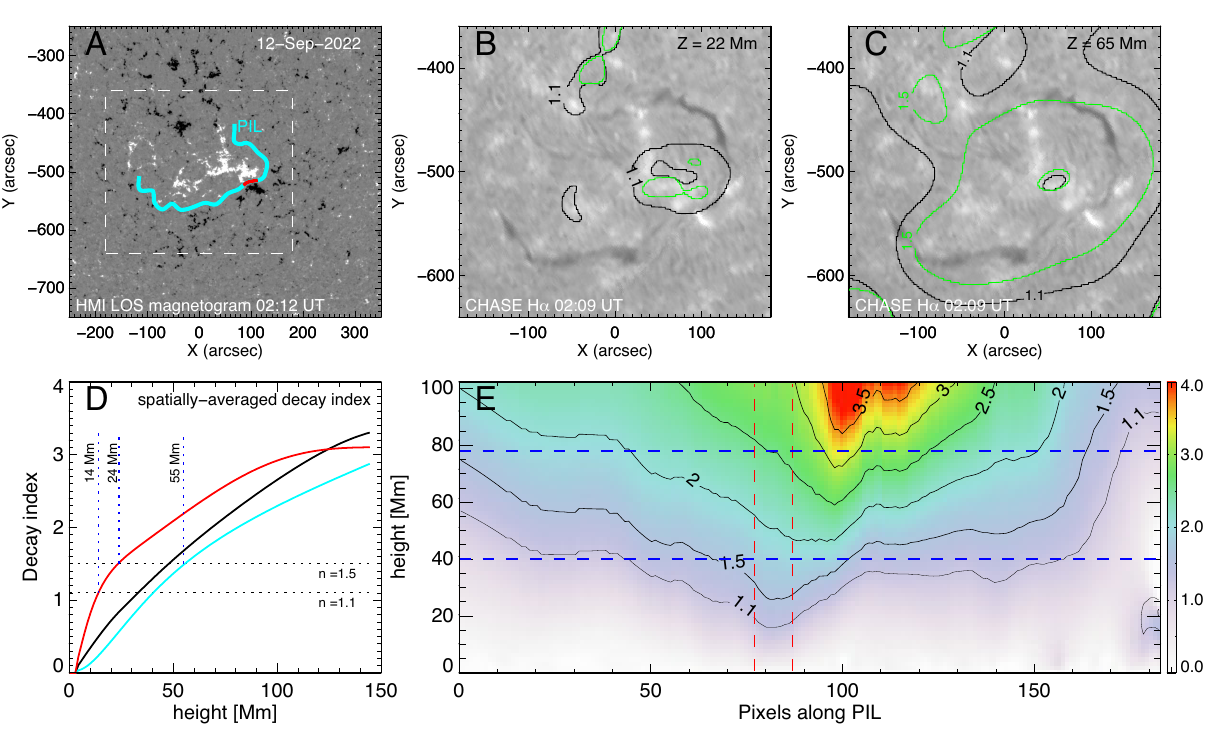}
\caption{Magnetic decay index analysis above the PIL of miniature and giant filaments on 2022 September 12th. (A) Magnetic environment of the source region, as it passed the central meridian on 2022 Sep 12. Polarity inversion lines (PILs) of the whole filament channel and of the minifilaments are outlined by the cyan and red lines, respectively. (B) and (C): Two-dimensional spatial distributions of decay index at the heights of 22 Mm and 65 Mm are plotted over a simultaneously taken CHASE H$\alpha$ image, respectively. (D) Spatially-averaged decay index as a function of height, which respectively are calculated over the whole PIL, the minifilament-related PIL (red section in (A)), as well as the rest PIL (two cyan sections in (A)). (E) The height variation of decay index above the whole PIL at 02:12 UT, in which contours at different levels are also plotted. Two red dashed vertical lines highlight the corresponding spatial range right above the PIL of minifilaments. Two blue dashed horizontal lines indicate the initial and maximum heights of the pre-eruption giant filament. }
\label{fig4}
\end{figure}

Three-dimensional simulations of the flux-cancellation model have demonstrated that slow flux cancellation process is responsible for the buildup and ascent of the filament magnetic structure, and the onset of torus instability triggers the final eruption of an MFR at a critical height \citep{2010ApJ...708..314A}. Based on magnetic decay index analysis, we find that the decay of background magnetic field in the region above the PIL exhibits obvious spatial non-uniformity. The two-dimensional spatial distribution of decay index (Figures~\ref{fig4} (B and C)) reveals that a torus-unstable domain appears at a very low height of 22 Mm above the photosphere near the middle segment of the PIL (where minifilaments erupted), and it extends to almost the whole PIL at the height of 65 Mm above the photosphere. The vertical spatial distribution of decay index along the PIL (Figure~\ref{fig4} (E)) and line profiles of spatially averaged decay index (Figure~\ref{fig4} (D)) also shows that the background magnetic field quickly decreases with height and this trend is most prominent above the middle part of the PIL. This non-uniformity of background magnetic fields naturally allows minifilaments to experience the torus instability due to the appearance of a low-lying torus-unstable domain above the middle segment of the long PIL, while preventing the giant filament from suffering the torus instability due to a high-lying torus-stable domain. 
Similar non-uniform decay of background magnetic fields had also been used to explained the onset of partial filament eruption in \citep{2018ApJ...869...78C}.

Prior to eruptions, heights of the third minifilament and giant filament were measured as $\sim$12 Mm and 40 Mm, respectively (Figure~\ref{fig3}(D); also see Methods in Appendix \ref{sec:E}). The minifilaments are more likely supported by a short and straight flux-rope configuration, which may even require a torus instability critical threshold of $\sim$ 1.0$-$1.1, or even lower than unity \citep{2021MNRAS.503.3926F}. In this case, they may become torus unstable at an even lower height, approximately 14 Mm above the photosphere (see Figure~\ref{fig4} (D)). As a result of continuous flux convergence and cancellation at the middle segment of the long PIL, minifilaments are prone to losing their equilibrium at first. Subsequently, due to a trio of minifilament eruptions, the giant filament experiences sudden large-amplitude oscillations, propelling it to ascend to greater heights. Indeed, the 3D reconstruction result reveals that the giant filament attained a height of approximately 78 Mm at the onset time (around 19:54 UT). At this altitude, the decay index surpasses the critical threshold for torus instability, thereby allowing the sudden eruption of the giant filament. 

\subsection{Mechanism behind the physical coupling between miniature and giant filament eruptions}

\begin{figure}[ht!]
\plotone{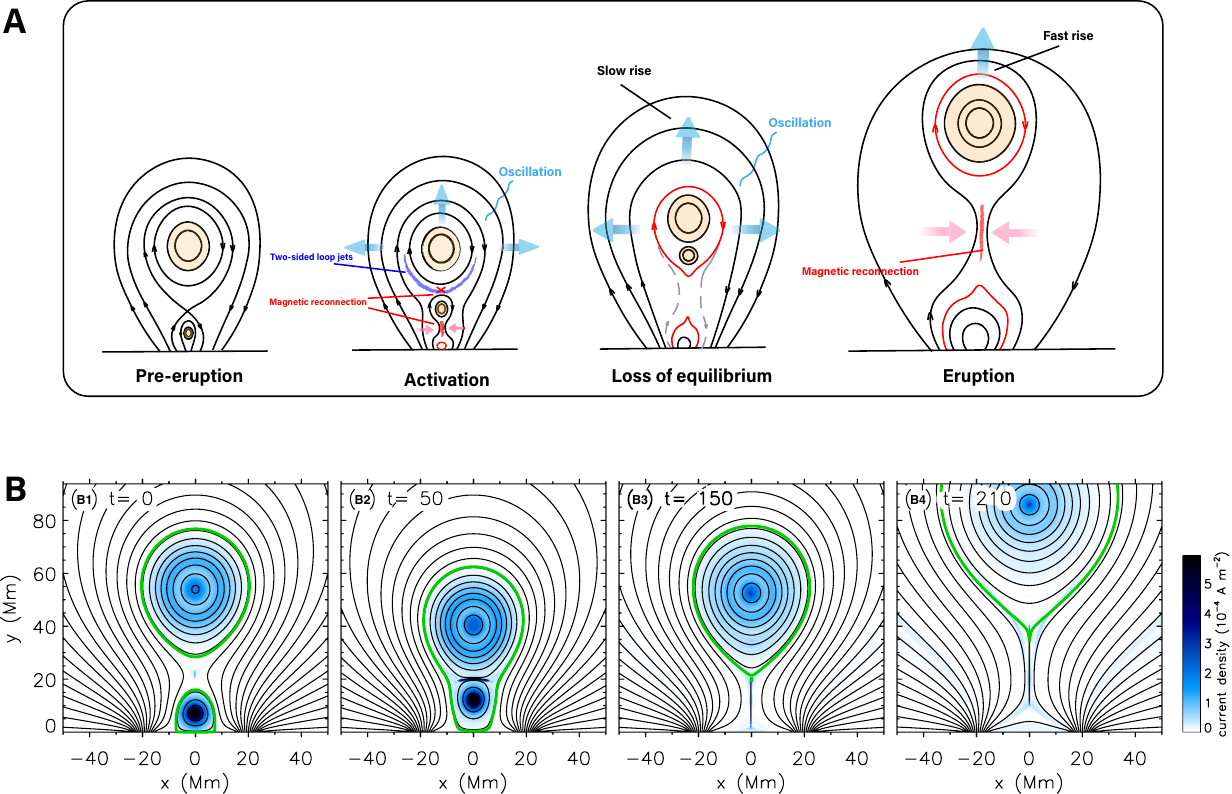}
\caption{Two-dimensional schematic view and 2.5-D MHD simulation of the coupling between miniature and giant filament eruptions. (A) In the pre-eruption phase, a mini-MFR and a giant MFR are both at the equilibrium state in the same filament channel, forming a double-decker system. In the activation phase, homologous mini MFRs sequentially lose equilibrium and erupt towards the overlying giant MFR, leading to the activation of the giant MFR. In this course, multiple magnetic interaction occurs between magnetic field of the erupting mini MFR and overlying giant MFR, including magnetic reconnection and flux feeding (or axial magnetic field merging), as well as oscillation. Reconnection occurs both ahead and behind the erupting mini MFR, which leads to the two-sided loop jets and tiny standard flare arcades, respectively. In the loss-of-equilibrium phase, the flux merging process between mini- and giant MFRs facilitates the disruption of the whole system. In the eruption phase, a newly formed large flux rope becomes globally unstable and steps into its final eruption. (B) 2.5-dimensional MHD simulation reproducing the interaction, flux merging, oscillation, and the whole eruption of this double-decker MFR system. (B1) An initial double-decker flux-rope system with an HFT configuration. (B2) The flux merging between a small and a large flux rope and subsequent flux rope oscillation. (B3) The ascent of the newborn flux rope; (B4) The eruption of the newborn flux rope. The times indicated in panels (B1)–(B4) are in unit of $\tau_A$. The blue color depicts the distribution of the current density. The green curves indicate the boundary of the big and small flux ropes.
An animation of Panel (B) is available in the online journal, depicting the 2.5D MHD simulation results. The duration of the animation is 13 s.
 }
\label{fig5}
\end{figure}

Combining the above observations and analysis results, we conclude that a series of minifilament eruptions driven by extremely weak flux cancellation break the initial equilibrium of an overlying giant filament via multiple interactions. We propose a scenario (depicted in Figure~\ref{fig5} (A)) to explain the unique physical coupling between miniature and giant filament eruptions. Initially, the coexistence of minifilament and giant filament, which are of the same magnetic helicity, within the same filament channel, forms a double-decker magnetic flux rope system \citep{2012ApJ...756...59L}. 
Due to the non-uniform decay of the background magnetic field, the high-lying giant filament remains in the torus-stable regime, while the background magnetic field above the minifilament decays rapidly, making it more prone to torus instability and eruption at a lower height. Under the continuous flux cancellation of small-scale magnetic fluxes at the middle segment of the PIL, minifilaments sequently form and erupt below the giant filament. The upward eruption of homologous minifilaments impacts the magnetic field at the bottom of the giant filament, triggering magnetic reconnection between the small and large flux ropes, as well as subsequent magnetic field merging between the flux ropes. Consequently, this reconnection process leads to the appearance of unique bidirectional coronal jets observed along the giant filament. Meanwhile, the erupting minifilament stretches the magnetic field upward, triggering subsequent standard flare reconnection and tiny two-ribbon flares. With multiple eruptions of minifilaments, similar processes repeat, and the merging of small and large magnetic flux ropes results in sustained disturbance and large-amplitude oscillations of the giant filament above. Eventually, the newly formed large flux rope becomes globally unstable and ascends into a higher torus-unstable regime, triggering a final global eruption and a two-ribbon flare at its source region. Recently, \citet{2024SoPh..299...85S} also reported a similar multi-filament eruption event, in which a large-scale filament eruption initiated by two preceding active-region filaments pushing out from below. However, the authors stated that there was no strong evidence of magnetic reconnection between the different filaments.  In contrast, our observations and magnetic field analysis provide a more comprehensive and clear insight into the formation and eruption of minifilaments, and the subsequent magnetic interaction between the mini- and giant filaments.

By employing 2.5-D magnetohydrodynamic (MHD) numerical simulations (Methods in Appendix \ref{sec:F}), we have successfully reproduced the key physical processes observed in this scenario. The simulation results in (see Figure~\ref{fig5} (B) and associated movie) are consistent with our observational facts, illustrating that the small flux rope undergoes initial instability and ascends within a double-decker magnetic rope system. Subsequently, upon collision with the larger flux rope above, reconnection and magnetic field merging take place, resulting in the formation of a newly created large flux rope with an increased axial flux. (Note that in 2.5-D simulations, magnetic field merging we observed often manifests as flux merging; therefore, the term ``flux merging" refers to the same process in our discussion here and in Appendix C.)
During this process, the newly formed large flux rope exhibits upward and downward oscillations. As the confinement of the background magnetic field remains unchanged, the augmented axial electric current within the large flux rope rapidly propels it upwards, abruptly leading to its disruption from a quasi-static equilibrium due to flux imbalance \citep{2020ApJ...898L..12Z}. This rising large flux rope then steps into its standard eruption, leaving a flare reconnection current sheet and associated post-flare loops behind.

 \section{Discussion}
 
 In the flux-cancellation model \citep{1989ApJ...343..971V}, magnetic flux cancellations provide a necessary environment for the buildup of filament magnetic fields. In this model, flux cancelation is suggested as the submerging of field lines after magnetic reconnection between the adjacent opposite-polarity feet of different magnetic arcades above the PIL. This reconnection-cancellation process would gradually create a newborn MFR with increasing helical field lines and also reduce the overlying confinement. As such a process proceeds, the newborn MFR would meet a series of quasi-static equilibria at progressively higher altitudes. Eventually, it would reach a critical point of force balance, where the whole filament magnetic structure experiences a catastrophic loss of equilibrium and inevitably erupts. Until now, almost all of leading theories \citep[e.g.,][]{1999ApJ...510..485A,2000JGR...105.2375L,2003ApJ...585.1073A,2010ApJ...708..314A,2021NatAs...5.1126J} and observational works \citep[e.g.,][]{2001ApJ...548L..99Z,2017ApJ...839..128W,2020A&A...642A.199Z,2021ApJ...919...34Y}  have exclusively considered the formation and eruption of isolated filaments. In particular, numerical simulations that related to flux cancellation have simply assumed that they are primarily governed by an idealized, uniform flux cancellation process \citep[e.g., ][]{2000ApJ...539..983V,2003ApJ...595.1231A,2010ApJ...708..314A,2014ApJ...780..130X}. However, the real flux cancellation is more complex than an idealized process and their potential impacts on filament activity have been largely overlooked in the past. These subjective assumptions also hinder us to uncover the potential physical connections between filament eruptions across different spatial-temporal scales. 

Our finding of a physical coupling between miniature and giant filament eruptions within the same filament channel provides us a unique perspective to reconsider the traditional flux cancellation model of solar eruptions. The HMI observations reveal that from September 8 to 16, prolonged flux cancellation along the PIL exhibits significant temporal and spatial complexity that far exceeds the assumptions of an ideal uniform flux cancellation process commonly considered in most existing simulations and theories of solar eruptions. The key aspect of this complex flux cancellation is the gradual reduction in the size of flux cancellation region over time, from a scale exceeding 100 Mm to a few Mm. Incorporating this fact allows for a more self-consistent understanding of all the reported phenomena above. First, this provides insight into why the giant filament and a series of minifilaments appear along the same PIL. Initially, flux cancellation at the scale of active regions establishes the large-scale filament channel and gives rise to the formation of the giant filament. Subsequently, the spatial region of flux cancellation gradually contracts to a few Mm, leading to the sequential formation of multiple minifilaments. Similar small-scale flux cancellation have been previously reported to be important for the formation and eruption of minifilaments at the base of coronal jets \citep{2017ApJ...844..131P,2018ApJ...853..189P,2020ApJ...902....8C}.
Second, this offers an explanation for the spatial non-uniformity in the decay of the background magnetic field above the PIL. Flux cancellation is closely associated with horizontal converging flows on the photosphere. Persistent converging flows are thought to be important in causing the gradual approaching of the photospheric footpoints of the background magnetic field, resulting in the elevation of the loop top of background magnetic field above the PIL \citep{1991ApJ...373..294F,1993ApJ...417..368I}. In the later stages of the observations, the continuous flux convergence and cancellation primarily occurs in the central region of the PIL. As a result, the decay of the background magnetic field along the PIL is most pronounced in the central portion, which provides a potential condition for the preferential eruption of minifilaments. Finally, this unveils a novel pathway through which the gradual evolution of the photospheric magnetic field initiates the abrupt eruption of large-scale filaments in the solar corona. The eruptions of minifilaments, driven by extremely weak flux cancellation within a spatial region of a few Mm, unexpectedly provide a novel avenue for small-scale magnetic activities at the solar surface to trigger the destabilization and eruption of the overlying giant filament.

In fact, there are some clues that the occurrence of miniature and giant filaments within a single filament channel may not be rare events \citep{2016RAA....16....3Y,2019ApJ...883..104S,2019ApJ...887..220Y,2019ApJ...871..220S,2024ApJ...964....7Y,2020MNRAS.498L.104W,2024ApJ...970..100T}, but the magnetic coupling of minifilaments to overlying giant filaments has been largely overlooked and solid evidence for their potential role in the triggering of solar eruptions has never been reported. A previous publication \citep{2014ApJ...789..133Z} has reported a prominence eruption driven by flux feeding from chromospheric fibrils, in which the flux feeding is quite similar to the horizontal flux merging reported here. The authors tended to attribute the origin of chromospheric fibrils to flux emergence below the overlying prominence, but the extra flux emergence is not inherent to the formation of overlying solar prominences. 
In the current work, we first propose that both minifilaments and the overlying filament are homologous solar magnetic activities driven by the same physical process but only occur at different tempo-spatial scales. Within this unified framework, the formation and eruption of minifilaments and the larger overlying filament can be interpreted as self-similar MHD phenomena driven by a prolonged flux cancellation process. Meanwhile, the continuous transfer of mass, horizontal magnetic flux, as well as momentum from minifilament eruptions, which are driven by the gradual evolution of the photospheric magnetic field, represents a novel pathway to disrupt the force-free equilibrium of the overlying filament structure. Thus, our finding introduces a new and important ingredient to solar eruption models that was not encompassed within the traditional flux cancellation framework. 


\begin{acknowledgments}
The authors sincerely thank the referee for constructive suggestions.
This work is supported by the NSFC grant 12425301, 12103005, 12333009, 12073022, 12373115, 42188101, and 42174213 and the National Key R\&D Program of China (nos. 2021YFA0718600). H. C. C. and X. C. are also supported by the Yunnan Key Laboratory of Solar Physics and Space Science (YNSPCC202210, 202205AG070009) and the Yunnan Provincial Basic Research Project (202401CF070165). 
Q. H. Z. is also supported by Strategic Priority Research Program of the Chinese Academy of Sciences (XDB0560302). C. X. and Z. Y. H. are also supported by the Strategic Priority Research Program of the Chinese Academy of Sciences (XDB0560000). This work uses data from the CHASE mission supported by the China National Space Administration. SUTRI is a collaborative project conducted by the National Astronomical Observatories of CAS, Peking University, Tongji University, Xi’an Institute of Optics and Precision Mechanics of CAS, and the Innovation Academy for Microsatellites of CAS. AIA data and HMI Magnetic field data are courtesy of NASA/SDO, a mission of NASA’s Living With a Star Program. STEREO EUVI data are courtesy of the STEREO/SECCHI consortium.
\end{acknowledgments}

%






\appendix
\renewcommand{\thefigure}{A\arabic{figure}} 
\setcounter{figure}{0} 

\section{Filament chirality and magnetic-field configuration} \label{sec:C}

\begin{figure}[ht!]
\plotone{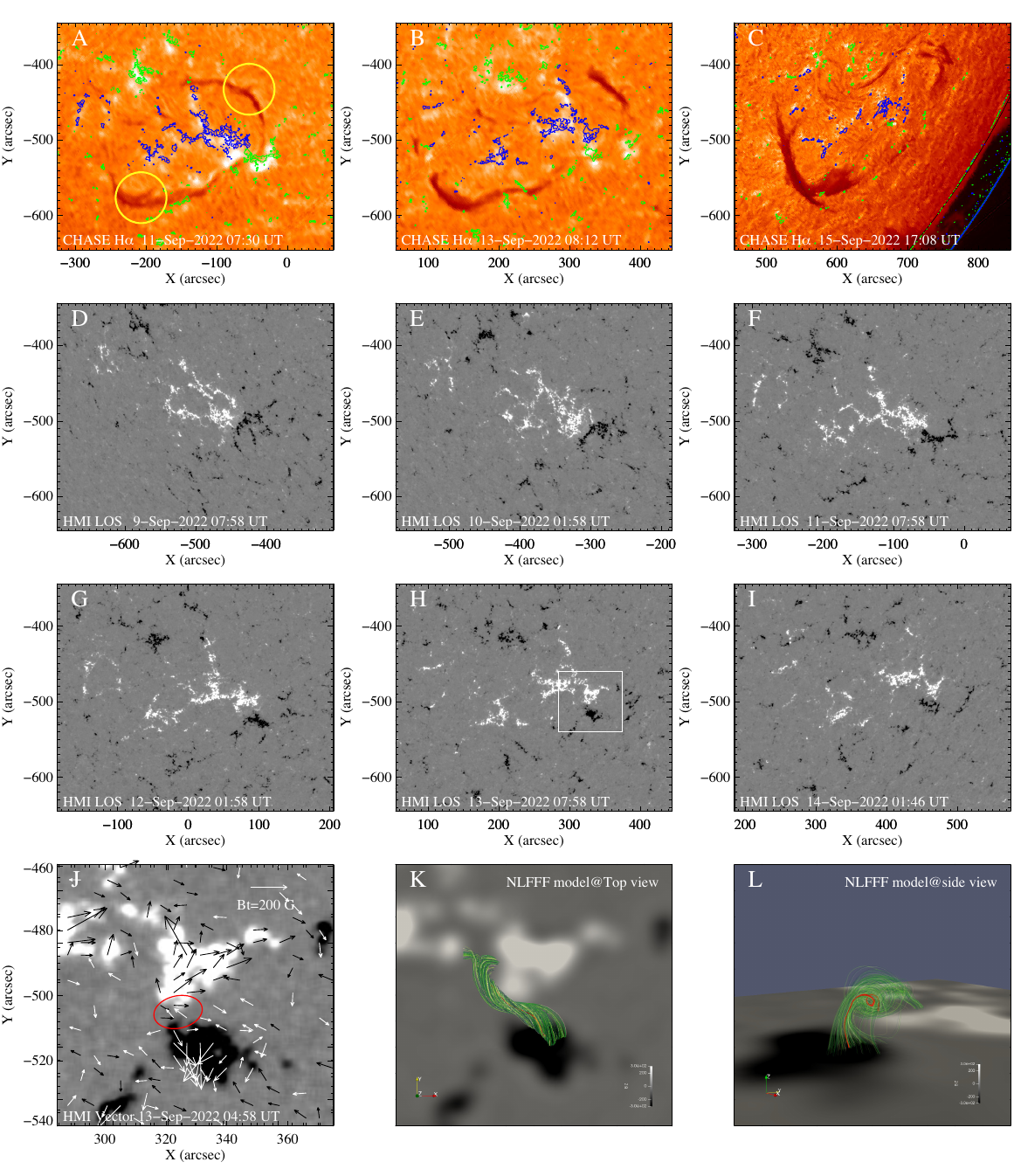}
\caption{The long-term evolution of the filament channel, where mini- and giant filaments later erupt. (A-C) Selected H$\alpha$  images taken by the CHASE, in which contours of simultaneous HMI magnetograms are overplotted at the level of ± 300 Gauss. (D-I) HMI LOS magnetograms from 9 Sep 2022 to 14 Sep 2022. (J) HMI vector magnetogram in the region where minifilaments later form and erupt (see white rectangle in (H)). The black and white arrows represent the strength of positive and negative horizontal magnetic field, respetively. Red ellipse highlights one plausible bold patch in this region. (K) Top view and (L) side view of the NLFFF model over this region. The background is HMI magnetograms with the level of ± 300 Gauss. In panel (L), one red line highlights the helical trend of field lines.}
\label{figA1}
\end{figure}
The giant filament was bred in a highly decayed active region, in which its photospheric vector magnetograms are too weak and descretized to perform a reliable coronal magnetic field extrapolation. Therefore, we judged the possible magnetic field configuration of the giant filament following an indirect method proposed based on the chirality and magnetic theoretical model of filaments \citep{2010ApJ...714..343G,2014ApJ...784...50C}. This indirect method suggests that “if a sinistral filament has left-bearing (or right-bearing) barbs and right-skewed (or left-skewed) overlying arcades, it should be supported in an inverse-polarity magnetic flux rope (MFR) configuration (or a normal-polarity sheared arcade configuration); if a dextral filament, the correspondence is the opposite." This method has been widely used to determine the possible magnetic topologies of various types of filaments/prominences \citep[e.g.,][]{2017ApJ...835...94O,2018ApJ...869...78C}. 
According to this indirect method, we can conclude that the giant filament exhibits sinistral characteristics and is supported by an inverse-polarity MFR. This conclusion is supported by several observational facts. First, when viewed from the positive-polarity side, the axial component field of the filament turns to the left, indicating its sinistral nature. Second, the giant filament displays left-bearing barbs, as depicted by the circles in Figure~\ref{figA1} (A). Furthermore, during the eruption of the filament, two conjugate footpoints of its drainage were identified in the CHASE doppler map, as shown by the arrows in Figure~\ref{fig2} (B). It is evident that the two draining sites of the giant filament exhibit a right skewness of strapping arcades. 

The minifilaments located below the giant filament, particularly the third one (Figure~\ref{fig1} (H)), exhibit a distinct “S" shape, which suggests that they also possess a sinistral chirality \citep{1998ASPC..150..419M}. We examined the HMI vector magnetograms in the source region of minifilaments as it passed the center of the solar disk and found a notably higher average horizontal magnetic field compared to other sections of the giant filament’s PIL. This thus allows a further analysis of magnetic topology and a possible magnetic field extrapolation in the source region of minifilaments. Fortunately, our observations indeed revealed the presence of bald patches above the forming site of the minifilament. This is clearly indicated by the red ellipse in Figure~\ref{figA1} (J), where an orientation reversal of the horizontal magnetic field occurs at the photosphere  \citep{1993A&A...276..564T}. This reversal indicates that helical field lines are tangent from above to the photosphere in that region. Moreover, using this vector magnetograms as boundary condition, the coronal magnetic field in the source region of minifilaments is extrapolated with the help of a nonlinear force-free extrapolation code provided in the SSWIDL package. This code is developed based on an optimization method \citep{2000ApJ...540.1150W,2004SoPh..219...87W,2012SoPh..281...37W}. Our results of the nonlinear force-free extrapolation, as depicted in Figure~\ref{figA1} (K and L), further demonstrate the existence of a small-scale MFR with sinistral chirality above the forming site of the minifilaments. These results suggest that these minifilaments are associated with MFR configurations, which is in line with previous arguments on minifilament observations from \citep{2020ApJ...902....8C}.

\section{3D reconstruction and the fitting of height-time profiles} \label{sec:E}

\begin{figure}[ht!]
\plotone{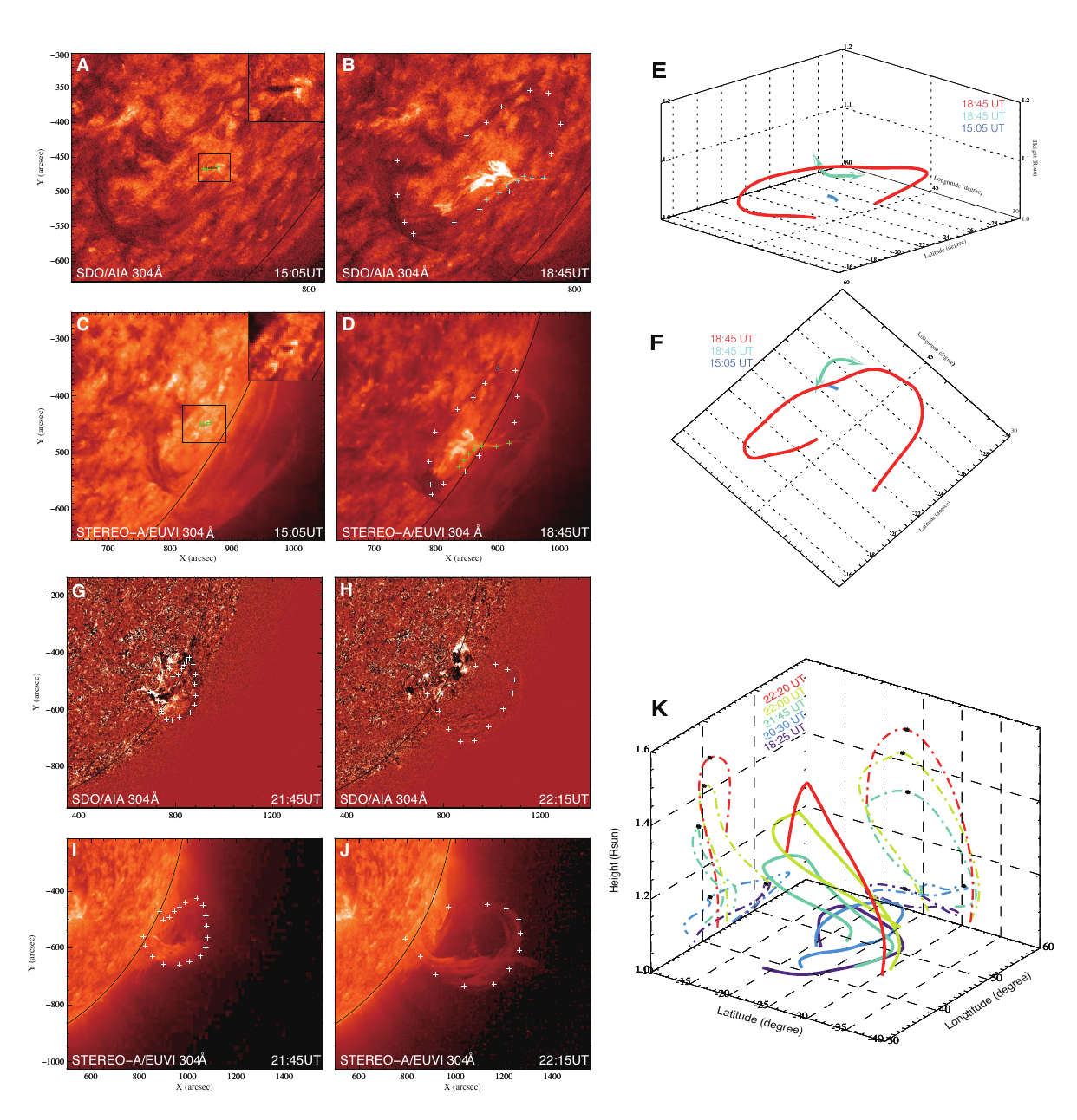}
\caption{Three-dimensional reconstruction from dual-perspective SDO/AIA and STEREO-A/EUVI observations. (A-D) Dual-perspective snapshots of the third minifilament and the pre-eruption giant filament, as well as the two-side loop jets. A closed-up view of the minifilament is shown at the right-up corner of (A) and (C), respectively. (E-F) 3D reconstructed morphology of the pre- eruption giant (red curve) and miniature filaments (blue curve), as well as two-side loop jets (cyan curve), viewed from two different perspectives. (G-J) Dual-perspective snapshots of the erupting giant filament at two different times. (K) 3D reconstruction results of the erupting giant filament at different times, and their projections on two different 2D planes. The black asterisks indicate the top of filament main axis, respectively.}
\label{figA6}
\end{figure}

The dual-perspective EUV imaging observations provided by the STEREO-A/EUVI and SDO/AIA enable us to reconstruct the 3D topology and height of pre-eruption filaments and the two-sided loop jets, as shown in Figure~\ref{figA6}. We chose the EUVI 304 images for 3D reconstruction. On 2022 September 15, the separation angle between STEREO-A and SDO was about 20°.  The 3D reconstruction was conducted using the  procedure \textit{scc$\_$measure.pro} in the SSWIDL package,  which is developed by \citet{2012SoPh..276..241T} and have been widely used in observational studies of solar filament eruptions \citep[e.g.,][]{2019ApJ...883..104S,2021ApJ...911...33C,2024ApJ...968..110D}. The procedure workflow is as follows: for each simultaneous time frame, two 304 \AA~images selected from STEREO-A/EUVI and SDO/AIA are displayed side by side, with a slight offset to account for the difference in light travel time from the Sun. The widget allows the user to zoom in on the region of interest in the two images, and optimize the appearance of the feature being measured though adjusting the color table or data range. The user then can select a common feature appearing in both images with the cursor. Once a point is selected in the first image, the procedure would calculate the three-dimensional epipolar line of sight of this selected point and then overplots the projection of this line onto the second image as a reference line.  Along the reference line, the user can easily pinpoint the same feature in the second image with the cursor, which leads to another line of sight calculation. As a result, the intersection of these two three-dimensional epipolar lines determines the 3D location of the feature. For example, for the giant filament, the height of the top of the filament axis was measured over time (Figure~\ref{figA6} (G to K)). Similarly, regarding the minifilament observed at 15:05 UT, its filament axis was reconstructed to estimate the approximate height of the homologous minifilament erupted at 18:05 UT; regarding the two-sided loop jet, its height and space trajectory was reconstructed in  Figure~\ref{figA6} (E and F).

\begin{figure}[ht!]
\plotone{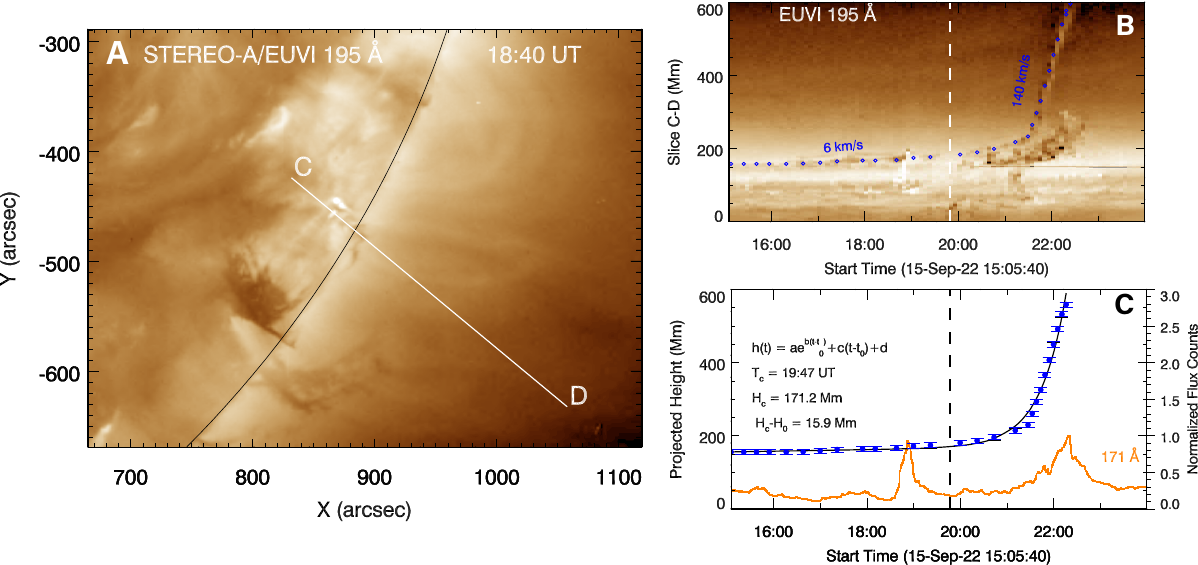}
\caption{The eruption of the giant filament and its height-time evolution observed from the perspective of STEREO-A on 15 September 2022. (A) The STEREO-A/EUVI 195 Å image at 18:40 UT. The Slice CD represents the radial eruption path of the giant filament. (B) The space-time diagram along slice CD, showing the height-time evolution of the giant filament eruption. (C) The height-time profile of the erupting giant filament and its fitting result. The AIA 171 Å flux variation computed in the source region (see the same FOV of Fig. 3E) is also plotted.
An animation of this Figure is available in the online journal. The animation begins before the time sequence shown in panel (A), with the EUVI 195 \AA~images (panel (A)) spanning from 15 September 2022 15:00 UT to 23:55 UT. The duration of the animation is 4 s.
}
\label{figA7}
\end{figure}

Figure \ref{figA7} presents the eruption of the giant filament and its height-time evolution observed from the perspective of STEREO-A.
The height-time profiles that we obtained from 3D reconstruction technique or time-distance diagram were well fitted by a function consisting of linear and exponential components \citep{2013ApJ...769L..25C}. This function can be defined as follows: \textit{h(t) = ae$^{b(t-t_0)}$+c(t-t$_0$)+d}, where \textit{h(t)} is height, \textit{t} is time, and \textit{a}, \textit{b}, \textit{c}, \textit{d}, and  \textit{t$_0$} are five free parameters. Using the fitting technique, the onset time (\textit{t$_c$} =$\frac{1}{b}ln(\frac{c}{ab})$+t$_0$ ) of the exponential acceleration of the giant filament eruption can be derived and its uncertainty can be estimated through Monte Carlo simulations. For each fitting, a critical height ($H_c$ =\textit {h(t$_c$})) , where the impulsive acceleration begins, can also be determined from this fitting technique. In our study, this fitting technique was applied in Figures~\ref{fig3}(E) and \ref{figA7}. 

\section{Magnetic field decay index analysis} \label{sec:F}
 The occurrence of ideal MHD torus instability may lead to the rapid acceleration of the filament-hosting MFR. The torus instability only takes place when the filament-hosting MFR ascends into a torus-unstable domain, where the decay of background strapping magnetic fields exceeds a threshold so that strapping fields are insufficient to balance the upward hoop force. The decay of background magnetic fields can be characterized by magnetic decay index. The magnetic decay index is calculated by the equation: $n = -\partial ln(B_h) / \partial ln(h)$\citep{2006PhRvL..96y5002K}, where $B_h$ represents the horizontal component of background potential magnetic field, and $h$ represents the height above the photosphere.  For a semi-circular flux tube, its critical theoretical threshold is about 1.3$-$1.5, while for a straight flux tube, its critical theoretical threshold is about 1.0$-$1.1 (see ref. \citep{2021MNRAS.503.3926F} and references therein). In the current work, we performed a potential field extrapolation for the whole source region of the giant and miniature filaments. 
While potential field extrapolation may not accurately capture the intricate magnetic structures of the filaments themselves, it has been proven to provide a reliable approximation to the overlying background magnetic field. Following the analysis method in \citep{2018ApJ...869...78C}, in the current work, both the spatial distribution and spatial-averaged profile of magnetic decay index were used to investigate the non-uniform decay of background magnetic field above the PIL. 
The analysis results of magnetic field decay index, calculated from magnetic field extrapolations on September 12 and 14, respectively. The results are generally consistent, but the magnetic field measurements on September 14 are subject to a more significant projection effect. Due to article length restrictions, only the results from September 12 are presented in Figure~\ref{fig4}.

\section{2.5-dimensional magnetohydrodynamic simulation} \label{sec:G}

Our 2.5-dimensional MHD simulation are conducted following the basic settings and similar parameters as in \citep{2021A&A...647A.171Z}. The magnetic field is described by the equation: $B = \nabla\psi \times \hat{z} + B_z \hat{z}$, where $\psi$ denotes the magnetic flux function and $B_z$ represents the component of the magnetic field in the z-direction. All the magnitudes  satisfy $\frac{\partial}{\partial z} = 0$, so that the evolution of magnetic field lines can be visualized in the x-y plane as isolines of $\psi$. The numerical domain is $0 <  x  < 200$ Mm, $0 <  y  < 300$ Mm, in which a symmetric condition is used in its left boundary ($x = 0$); increment equivalent extrapolation is used at the other boundaries \citep{2023FrASS...984678Z}. The background field is open bipolar field with $B_z=0$ , whose magnetic charge are located at the lower photosphere within $-25<$ x $<-15$ Mm and $15<$ x $<25$ Mm, respectively. An anomalous resistivity is used in the simulation to restrict magnetic reconnection within the regions of current sheets:

\begin{equation}
\eta = 
\begin{cases}

    0, & \text{if } j < j_c, \\
    (\eta_m \mu_0 v_0 L_0) \left(\frac{j}{j_c} - 1\right)^2, & \text{if } j > j_c.
\end{cases}
\end{equation}

Here \( L_0 = 10^{-7} \, \text{m}, \eta_m = 0.10, v_0 = 128.57 \, \text{km s}^{-1},  j_c = 4.45 \times 10^{-4} \, \text{A m}^{-2} \) is the critical current density, and \( \mu_0 \) is the vacuum magnetic permeability. Starting from the initial state described in \citep{2021A&A...647A.171Z}, we initiated the emergence of a small flux rope from the lower boundary of the domain using procedures similar to those outlined in \citep{2021A&A...647A.171Z}; the magnetic field strength within the emerging small rope is carefully tuned to be sufficiently weak, so that the small rope does not rise and merge with the pre-existing coronal flux rope but sticks to the photosphere after the emergence. The magnetic fluxes of the two flux ropes are then adjusted by similar procedures as those in \citep{2003JGRA..108.1072H}, so as to achieve an equilibrium state, as illustrated in Figure~\ref{fig5} (b1), where a minor/major flux rope is situated below/above the X-type magnetic structure, creating a double-decker flux rope configuration.  This equilibrium state serves as the initial condition for our simulation. The azimuthal and axial flux of the major flux rope are set at $\Phi_{p0}^{m}=11.2\times10^9$ Mx cm$^{-1}$ and $\Phi_{z0}^{m}=48.5\times10^{18}$ Mx, respectively, while the azimuthal and axial flux of the small rope are $\Phi_{p0}^{s}=5.6\times10^9$ Mx cm$^{-1}$ and $\Phi_{z0}^{s}=6.3\times10^{18}$ Mx, respectively. With these initial condition and the background configuration, numerical simulations are conducted to investigate the evolution of this double-decker flux rope system.

In our simulation, the magnetic fluxes of the major flux rope remain constant at $\Phi_{p0}^{m}$ and $\Phi_{z0}^{m}$, while the azimuthal flux of the small flux rope is also held fixed at $\Phi_{p0}^{s}$. In order to induce the subsequent eruption of the small flux rope, we implement a linear increase in its axial flux from $\Phi_{z0}^{s}$ to $6.7\times10^{18}$ Mx over the period 0-5$\tau_A$, where $\tau_A = 17.4$ s represents the characteristic Alfvén transit time. After 5$\tau_A$, all the magnetic fluxes of the small and major ropes remain constant, allowing the system to evolve in a relaxed manner. As illustrated in Figure~\ref{fig5}, the small flux rope gradually ascends and interacts with the major flux rope. With the ascent of the small flux rope, the major flux rope quickly undergoes a downward motion. By t = 50$\tau_A$, a horizontal current sheet emerges between the boundaries of the small and major flux ropes, where the current density distribution is depicted in blue (Figure~\ref{fig5} (B2)). Consequently, the two flux ropes undergo reconnection and merge to form a larger newborn flux rope (Figure~\ref{fig5} (B3)). This newly formed large flux rope ascends, creating a vertical current sheet in its path, and eventually erupts, as demonstrated in Figure~\ref{fig5} (B3)-(B4).

\bibliography{sample}\bibliographystyle{aasjournal}



\end{document}